\documentclass[preprint,12pt]{elsarticle}
\usepackage{graphicx}

\begin{document}

\centerline{\Large $\Lambda NN$ input to neutron stars from hypernuclear data}

\centerline{E. Friedman, A. Gal,}
\centerline{Racah Institute of Physics,  Hebrew University, Jerusalem 9190400, Israel}
\vspace{3mm}

This work is a sequel to our two 2023 publications [PLB 837 137669, 
NPA 1039 122725] where fitting 14 1$s_\Lambda$ and 1$p_\Lambda$ 
single-particle binding energies in hypernuclei across the periodic table led 
to a well-defined $\Lambda$-nucleus optical potential. The potential consists 
of a Pauli modified linear-density ($\Lambda N$) and a quadratic-density 
($\Lambda NN$) terms. The present work reports on extending the above analysis 
to 21 $\Lambda$ single-particle data points input by including 1$d_\Lambda$ 
and 1$f_\Lambda$ states in medium-weight and heavy hypernuclei. 
The upgraded results for the $\Lambda N$ and $\Lambda NN$ 
potential depths at nuclear-matter density $\rho_0=0.17$~fm$^{-3}$, 
$D^{(2)}_\Lambda=-37.5\mp 0.7$~MeV and $D^{(3)}_\Lambda=+9.8\pm 1.2$~MeV 
together with the total depth $D_\Lambda=-27.7\pm 0.5$~MeV, agree within 
errors with the earlier results. The $\Lambda$ hypernuclear overbinding 
associated with the $\Lambda N$-induced potential depth $D^{(2)}_\Lambda$ 
agrees quantitatively with a recent combined analysis of low-energy 
$\Lambda p$ scattering data and correlation functions [PLB 850 (2024) 138550]. 
These results, particularly the size of the repulsive $D^{(3)}_\Lambda$, 
provide an essential input towards resolving the 'hyperon puzzle' in the core 
of neutron stars. We also show that a key property of our $\Lambda NN$-induced 
potential term, i.e. a need to suppress the quadratic-density $\Lambda NN$ 
term involving an excess neutron and a $N=Z$ core nucleon, can be tested in 
the forthcoming JLab E12-15-008 experiment. 

\vspace{3mm}
E-mail: eliahu.friedman@mail.huji.ac.il; avraham.gal@mail.huji.ac.il

\vspace{3mm}
\noindent Submitted to

\noindent  Proceedings of {International Conference on Exotic Atoms and Related 
\newline Topics 
and Conference on Low Energy Antiprotons (EXA-LEAP2024),\\
 26-30 August 2024,
Austrian Academy of Sciences, Vienna.\\}

\newpage

\section{Introduction} 
\label{sec:intro}

Binding energies of $\Lambda$ hyperons in single-particle states of 
$\Lambda$ hypernuclei along the periodic table have been studied since the 
1970s~\cite{GHM16}. Although these data are fitted by several versions of 
optical-model and Skyrme-Hartree-Fock (SHF) approaches that differ in their 
high powers of the nuclear density $\rho$, the need to consider high powers of 
$\rho$ has not been demonstrated unambiguously. Recent observations of neutron 
stars (NS) with mass exceeding twice solar mass are in conflict with the 
expectation that $\Lambda$ hyperons in NS cores would soften the NS equation 
of state, thus preventing NS masses from exceeding $\sim$1.5 solar mass. 
However, a mechanism of inhibiting the appearance of $\Lambda$ hyperons 
through a repulsive $\Lambda NN$ interaction apparently explains the 
astrophysical observations~\cite{GKW20}. 

The present work offers a brief summary and an extension of our recent 
work~\cite{FGa23a,FGa23b} on this topic. Using a simple density-dependent 
optical potential scheme, we check whether a {\it repulsive} $\Lambda NN$ 
term is required by fits to the binding-energy data. In Sect.~\ref{sec:method} 
the $\Lambda$ optical potential is described focussing on nuclear densities 
based on nuclear sizes. Section~\ref{subsec:extra} shows optical-potential 
predictions obtained by fitting 1$s_\Lambda$ and 1$p_\Lambda$ binding energies 
in $^{16}_{~\Lambda}$N and then extrapolating the potential to heavier 
species. A need to suppress the $\Lambda NN$ term when one nucleon is in the 
charge-symmetric core and the other is an `excess' neutron is established, 
apparently related to its isospin dependence. Section~\ref{subsec:chi2} shows 
$\chi ^2$ fits to the full data set, including 1$d_\Lambda$ and 1$f_\Lambda$ 
states not shown yet. Section~\ref{subsec:iso} demonstrates a way to study 
the isospin dependence of the $\Lambda NN$ term in 
$^{40,48}$Ca($e,e'K^+)^{40,48}_{~~~~\Lambda}$K electroproduction experiments. 
And the last section presents comparisons with several interaction models in 
which the $\Lambda N$ interaction overbinds $\Lambda$ hypernuclei, thereby 
implying that the corresponding $\Lambda NN$ contribution is indeed 
{\it repulsive}.

\section{Methodology} 
\label{sec:method} 

The $\Lambda$-nuclear optical potential employed recently in our 2023 works 
\cite{FGa23a,FGa23b} is of the form 
$V_{\Lambda}^{\rm OPT}(\rho)=V_{\Lambda}^{(2)}(\rho)+V_{\Lambda}^{(3)}(\rho)$,
representing a two-body $\Lambda N$ interaction 
\begin{equation} 
V_{\Lambda}^{(2)}(\rho) = -\frac{4\pi}{2\mu_{\Lambda}}\frac{f^{(2)}_A\,b_0}
{1+\frac{3k_F}{2\pi}f^{(2)}_A\,b_0}\,\rho  
\label{eq:V2} 
\end{equation} 
and three-body $\Lambda NN$ interaction 
\begin{equation} 
V_{\Lambda}^{(3)}(\rho) = +\frac{4\pi}{2\mu_{\Lambda}}f^{(3)}_A\,B_0\,
\frac{\rho^2}{\rho_0}, 
\label{eq:V3} 
\end{equation}
with $b_0$ and $B_0$ strength parameters in units of fm ($\hbar=c=1$).
In these expressions, $\mu_{\Lambda}$ is the $\Lambda$-nucleus reduced mass, 
$\rho$ is nuclear density normalized to the mass number $A$ of the nuclear 
core, $\rho_0=0.17$~fm$^{-3}$ stands for nuclear-matter density and 
$f^{(2,3)}_A$ are kinematical factors involved in going from the $\Lambda N$ 
and $\Lambda NN$ c.m. systems, respectively, to the $\Lambda$-nucleus c.m. 
system~\cite{FGa07}: 
\begin{equation} 
f^{(2)}_A=1+\frac{A-1}{A}\frac{\mu_{\Lambda}}{m_N},\,\,\,\,\,\,
f^{(3)}_A=1+\frac{A-2}{A}\frac{\mu_{\Lambda}}{2m_N}. 
\label{eq:fA} 
\end{equation}
The novelty of this presentation is connected to including explicitly Pauli 
correlations through the dependence of $V^{(2)}$ on the Fermi momentum 
$k_F=(3{\pi^2}\rho/2)^{1/3}$ which affects strongly the balance between the 
derived values of potential depths $D_\Lambda^{(2)}$ and $D_\Lambda^{(3)}$. 
This form for including Pauli correlations was suggested in Ref.~\cite{WRW97} 
and practised in $K^-$ atoms studies~\cite{FGa17}. 
It is missing in most of SHF studies.
We remark that the 
$\Lambda NN$ potential term $V_{\Lambda}^{(3)}(\rho)$ derives mostly 
from OPE diagrams with $\Sigma NN$ and $\Sigma^\ast NN$ intermediate 
states~\cite{GSD71}. Purely $\Lambda NN$ contributions are generated by 
$NN$ short-range correlations~\cite{WRW97}, estimated in the present context 
to affect the derived value of $D_\Lambda^{(3)}$ by few percents at most. 
Finally, we note that the low-density limit of $V_{\Lambda}^{\rm OPT}$ 
requires that $b_0$ is identified with the c.m. $\Lambda N$ spin-averaged 
scattering length~\cite{FGa07}, taken positive here.

In optical-model applications similar to the one adopted here it is crucial 
to ensure that the radial extent of the densities, e.g., their r.m.s. radii, 
follow closely values derived from experiment. With $\rho(r)=\rho_p(r)+
\rho_n(r)$, the sum of proton and neutron density distributions, respectively, 
we relate the proton densities to the corresponding charge densities where
the finite size of the proton charge and recoil effects are included.
This approach is equivalent to assigning some finite range to the
$\Lambda$-nucleon interaction. For the lightest elements in the database, 
harmonic-oscillator type densities were used for protons, assuming that  
the same radial parameters apply also for the corresponding neutron 
densities. For species beyond the nuclear $1p$ shell we chose 
two-parameter or three-parameter Fermi distributions normalized to $Z$ for 
protons and $N=A-Z$ for neutrons, derived from nuclear charge distributions 
assembled in Ref.~\cite{AM13}. For medium-weight and heavy nuclei, the r.m.s. 
radii of neutron density distributions assume larger values than those 
for proton density distributions, as practised in analyses of exotic 
atoms~\cite{FGa07}. Once neutrons occupy single-nucleon orbits 
beyond those occupied by protons, it is useful to represent the nuclear 
density $\rho(r)$ by
\begin{equation}
\rho(r)=\rho_{\rm core}(r)+\rho_{\rm excess}(r),
\label{eq:exc1}
\end{equation}
where $\rho_{\rm core}$ refers to the $Z$ protons plus the charge-symmetric
$Z$ neutrons occupying the same nuclear `core' orbits, and $\rho_{\rm excess}$
refers to the $(N-Z)$ `excess' neutrons.

\section{Results}
\label{sec:results}

\subsection{Extrapolations from $^{16}_{~\Lambda}$N}
\label{subsec:extra}

\begin{figure}[!htb]
\begin{center}
\includegraphics[width=0.7\textwidth]{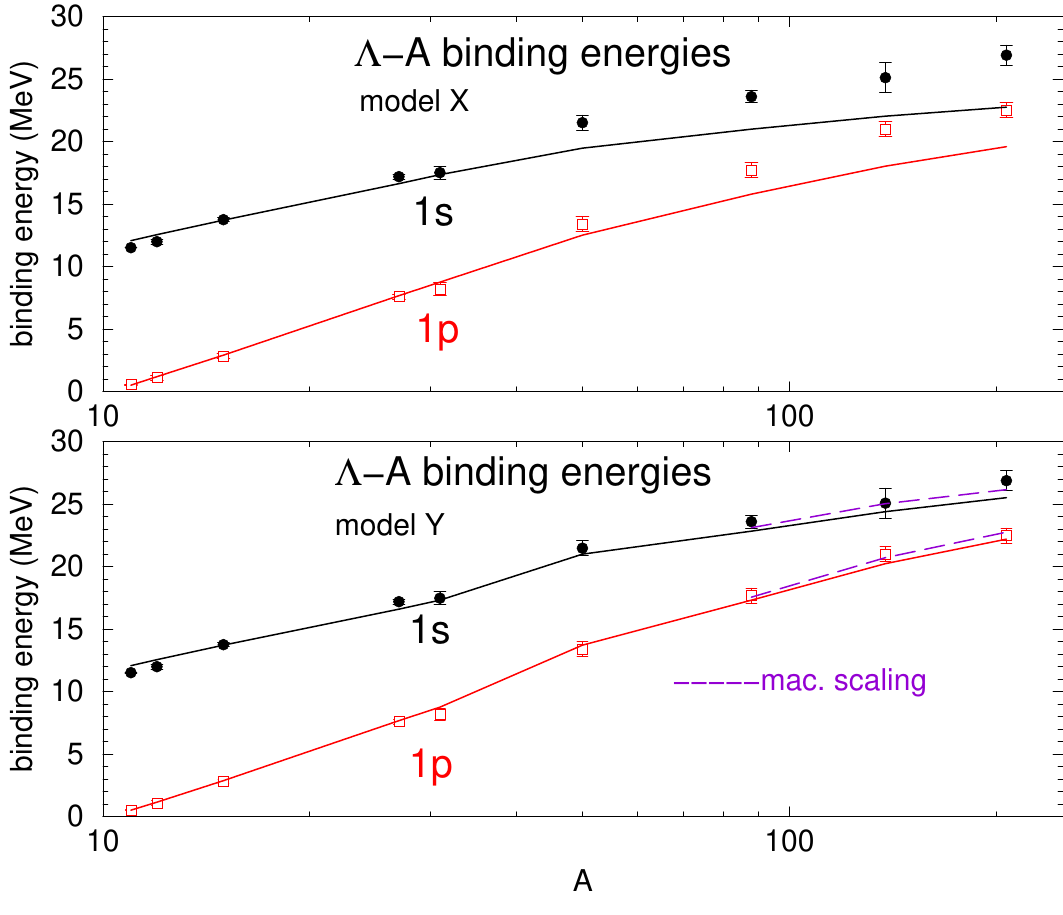}
\caption{$B_{\Lambda}^{1s,1p}(A)$ values across the periodic table as
calculated in models X (upper) and Y (lower), compared with data points,
including uncertainties. Continuous lines connect calculated values.
Figure updating Fig.~3 in Ref.~\cite{FGa23a}. The upper part, model X, 
uses the full $\rho^2$ term. The lower part, model Y, replaces $\rho^2$ 
by a reduced form, decoupling (N$-$Z) excess neutrons from $2Z$ symmetric-core 
nucleons, see text. The dashed lines are for $\rho^2$ replaced by $F\rho^2$, 
with a suppression factor $F$ given by Eq.~(\ref{eq:F}) below.}
\label{fig:EXA24_2}
\end{center}
\end{figure}

As a preliminary test of the optical-model approach we chose to fit the 
binding energies of 1$s_\Lambda$ and 1$p_\Lambda$ states in the hypernucleus 
$^{16}_{~\Lambda}$N because the simple 1$p$ proton-hole structure of
the $^{15}$N nuclear core removes, in this case, some of the uncertainty
from spin-dependent $\Lambda N$ and $\Lambda NN$ interactions. 1$s_\Lambda$ and 
1$p_\Lambda$ binding energies in $\Lambda$ hypernuclei, deduced from several 
strangeness production reactions, are listed in Table IV of Ref.~\cite{GHM16} 
including experimental uncertainties, and summarized 
in Table 2 of Ref.~\cite{FGa23b}.

The top part of Fig.~\ref{fig:EXA24_2} (model X) shows comparisons 
with experiment when the parameters $b_0$ and $B_0$ of 
Eqs.~(\ref{eq:V2},\ref{eq:V3}), determined by a fit to the two binding 
energies in $^{16}_{~\Lambda}$N, are used to {\it predict} 1$s_\Lambda$ 
and 1$p_\Lambda$ binding energies along the periodic table. Clearly seen 
is underbinding of 1$s_\Lambda$ and 1$p_\Lambda$ states in the heavier 
hypernuclei, argued~\cite{FGa23a,FGa23b} to result from treating equally all 
$NN$ pairs, including pairs where one nucleon is in the nuclear `core' while 
the other is an `excess' neutron. Removing the bilinear term from $\rho ^2$ 
using Eq.~(\ref{eq:exc1}) we replace $\rho ^2$ for heavy species by
\begin{equation}
\rho_{\rm core}^2+\rho_{\rm excess}^2\rightarrow(2\rho_p)^2+(\rho_n-\rho_p)^2,
\label{eq:exc2}
\end{equation}
in terms of the available densities $\rho_p$ and $\rho_n$. This prescription 
follows from the ${\vec\tau}_1\cdot{\vec\tau}_2$ isospin dependence arising 
from intermediate $\Sigma$ and $\Sigma^\ast$ hyperons in the $\Lambda NN$ OPE 
interaction~\cite{GSD71}. The reduced repulsion by the $\Lambda NN$ term is 
seen in the lower part of Fig.~\ref{fig:EXA24_2} (model Y). 

In the spirit of the present approach of avoiding explicit models, 
one can simply multiply $\rho^2$ by a suppression factor 
\begin{equation}
F=\frac{(2Z)^2+(N-Z)^2}{A^2},
\label{eq:F}
\end{equation}
that is approximately the ratio of the volume integral of
$(2\rho_p)^2+(\rho_n-\rho_p)^2$ to that of $\rho^2$. 
This ratio is significant, going down to $F$=0.67 for Pb. 
Also seen in the figure as dashed lines (`mac. scaling') are results 
using $F\rho^2$ instead of Eq.~(\ref{eq:exc2}), leading to almost identical 
results for the two options. It is interesting that extrapolating from 
$^{16}_{~\Lambda}$N to $^{208}_{~~\Lambda}$Pb is successful in describing 
the experimental results, while going down from $^{16}_{~\Lambda}$N into the 
nuclear $p$-shell (two smallest A-values) does not support the present simple 
model.

\subsection{Least-squares fits}
\label{subsec:chi2}

\begin{figure}[!ht]
\begin{center}
\includegraphics[height=7.0cm,width=0.70\textwidth]{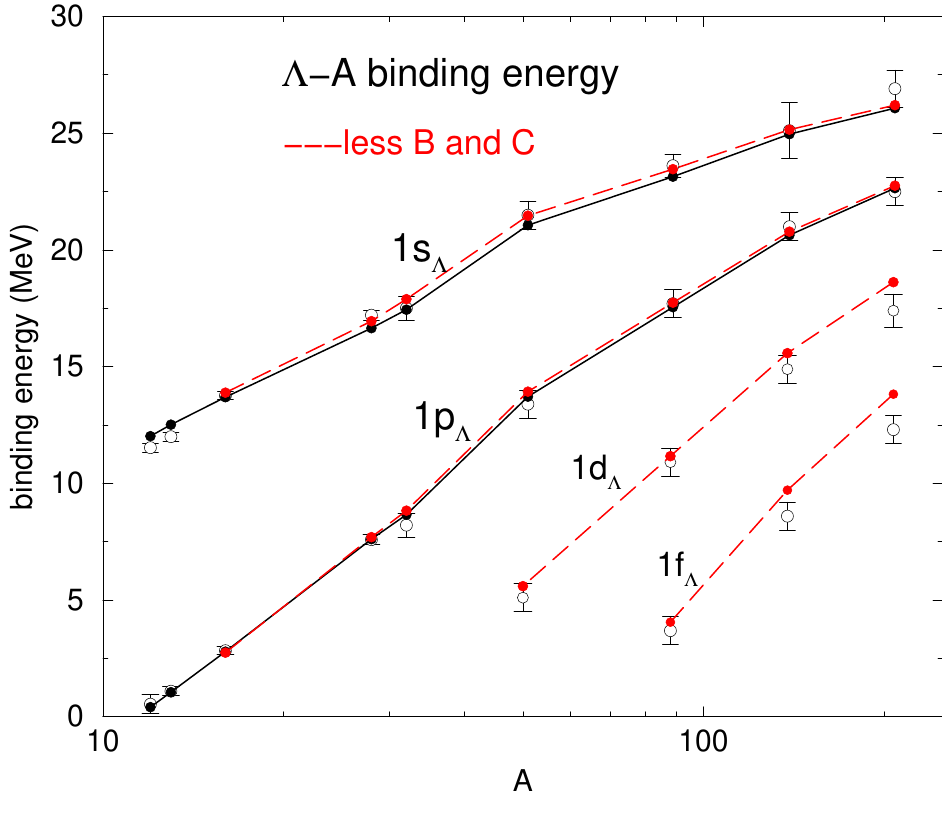}
\caption{$\chi^2$ fits to the full 1$s_\Lambda$ and $1p_\Lambda$ data 
(solid black lines) and when excluding $^{12}_{~\Lambda}$B and 
$^{13}_{~\Lambda}$C (dashed red lines). Also shown are {\it predictions} 
of 1$d_\Lambda$ and 1$f_\Lambda$ 
binding energies for the latter choice.}
\label{fig:LambdaALL}
\end{center}
\end{figure}

Next we performed conventional $\chi^2$ fits with the suppression 
factor $F$ included for the four heaviest species.
Figure \ref{fig:LambdaALL} shows several fits to the $B_{\Lambda}$ data
where black solid lines show fits to the full data set 
and open circles with error bars mark $B_{\Lambda}$ data points. It is
clearly seen that the $1s_{\Lambda}$ states in $^{12}_{~\Lambda}$B and
$^{13}_{~\Lambda}$C do not fit into the otherwise good agreement with
experiment for the heavier species. The red dashed lines show a very good fit
obtained upon excluding these light elements from the $B_{\Lambda}$ data set.
In fact, the potential parameters $b_0$ and $B_0$ of 
Eqs.~(\ref{eq:V2},\ref{eq:V3}) are hardly affected by the two lightest 
$^{12}_{~\Lambda}$B and $^{13}_{~\Lambda}$C hypernuclei. The fit parameters 
are:

{$b_0 = 1.44\,\pm\,0.095~{\rm fm}\,\,\ ({\rm attraction}),\,\,\,\, 
B_0 = 0.190\,\pm\,0.024~{\rm fm}\,\,\ ({\rm repulsion}),$} 
\newline 
with 100\% correlation between the two parameters. 
The fully correlated partial potential depths and the full one 
at nuclear-matter density $\rho_0=0.17$~fm$^{-3}$ are (in MeV):
\begin{equation} 
D^{(2)}_\Lambda=-38.6\mp 0.8,\,\,\,\,\,\,
D^{(3)}_\Lambda=11.3\pm 1.4,\,\,\,\,\,\,
D_\Lambda=-27.3\pm 0.6. 
\label{eq:total} 
\end{equation} 

Also included in Fig.~\ref{fig:LambdaALL} are {\it predictions} of 
1$d_\Lambda$ and 1$f_\Lambda$
binding energies made with these parameters. 
Although it is not expected that higher states will be well described by the 
same potential, owing to overlooked secondary effects such as non-local terms, 
it is seen that while slight overbinding of the calculated energies appears 
for the heavier species, the present optical potential reproduces quite well 
the four deepest $\Lambda$ single-particle states in neutron-rich hypernuclei. 
It is therefore of interest to repeat the $\chi^2$ process on the whole 
set of experimental binding energies of $\Lambda$ single-particle states, 
a total of 21 binding-energy values between $^{16}_{~\Lambda}$N and 
$^{208}_{~~\Lambda}$Pb. The resulting $\chi^2$ per degree of freedom is then 
0.95 (compared to 0.6 from a fit to only 14 binding-energy values for the 
1$s_\Lambda$ and 1$p_\Lambda$ states) 

{$b_0 = 1.32\,\pm\,0.072~{\rm fm}\,\,\, ({\rm attraction}),\,\,\,\,
B_0 = 0.162\,\pm\,0.020~{\rm fm}\,\,\, ({\rm repulsion}),$}
\newline
with 100\% correlation between the two parameters. 
The fully correlated partial potential depths and the full one 
at nuclear-matter density $\rho_0=0.17$~fm$^{-3}$ are (in MeV): 
\begin{equation}
D^{(2)}_\Lambda=-37.5\mp 0.7,\,\,\,\,\,\,
D^{(3)}_\Lambda=9.8\pm 1.2,\,\,\,\,\,\,
D_\Lambda=-27.7\pm 0.5. 
\label{eq:total2} 
\end{equation} 
These values are in agreement with those in Eq.~(\ref{eq:total}) based only on 
1$s_\Lambda$ and 1$p_\Lambda$ states. The uncertainties in the parameter 
values quoted above are statistical only. To estimate systematic effects 
within the adopted model we repeated the analysis with slightly modified 
nuclear densities such as obtained when unfolding the finite size of the 
proton. Values of $b_0$ came out unchanged whereas values of $B_0$ increased 
typically by 0.015~fm. The total potential depth at $\rho_0=0.17$~fm$^{-3}$ 
changed to $D_\Lambda=-26.9\pm 0.4$~MeV, suggesting a systematic uncertainty 
of 1 MeV for this value.

\subsection{Isospin dependence}
\label{subsec:iso}

\begin{figure}[!ht]
\begin{center}
\includegraphics[height=8.0cm,width=0.70\textwidth]{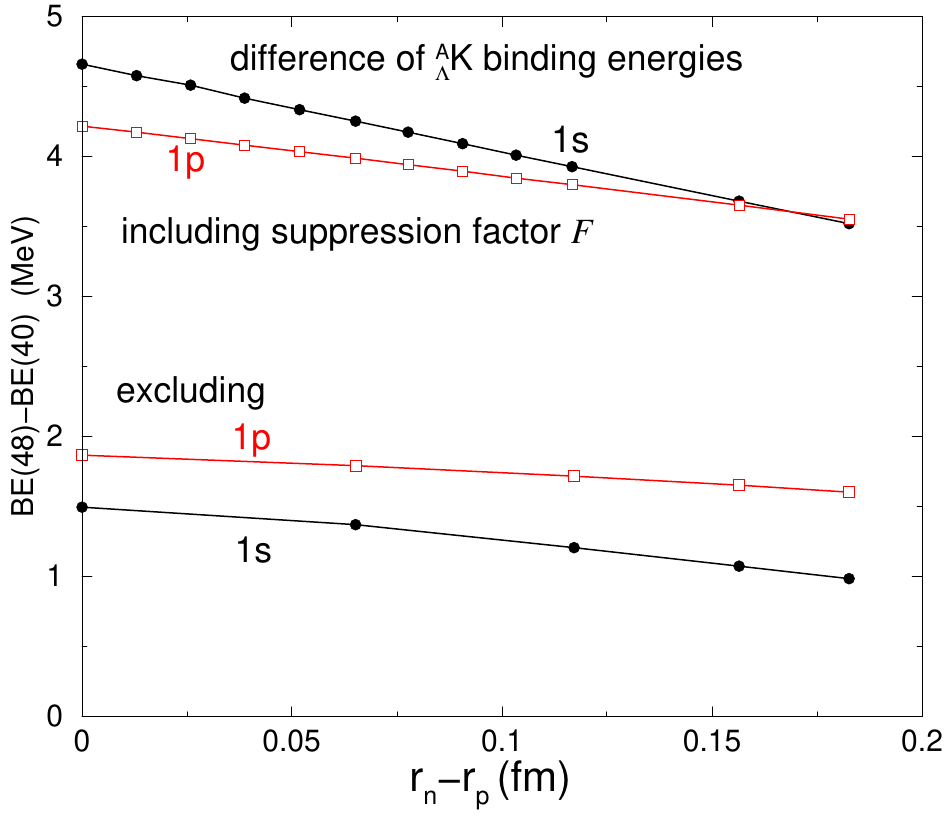}
\caption{$B_\Lambda(^{48}_{~\Lambda}{\rm K})-B_\Lambda(^{40}_{~\Lambda}{\rm K}
)$ values for 1$s_{\Lambda}$ and 1$p_{\Lambda}$ states, with and without 
applying the suppression factor $F$, as a function of the neutron-skin of 
$^{48}_{~\Lambda}$K, see text.}
\label{fig:dBvsdr2}
\end{center}
\end{figure}

Figure \ref{fig:dBvsdr2} shows calculated differences of $\Lambda$
binding energies for the $1s_{\Lambda}$ and $1p_{\Lambda}$ states between
$^{48}_{~\Lambda}$K and $^{40}_{~\Lambda}$K as a function of the neutron skin
$r_n-r_p$ in $^{48}_{~\Lambda}$K. The figure shows predictions made
using our standard $\Lambda$-nucleus potential $V_{\Lambda}^{\rm OPT}$ upon
including (excluding) in its upper (lower) part the suppression factor $F$, 
Eq.~(\ref{eq:F}). Regardless of the chosen value of $r_n-r_p$, the effect of 
applying $F$ is about 2.5~MeV for the $1s_{\Lambda}$ state and more than 2~MeV 
for the $1p_{\Lambda}$ state, within reach of the ($e,e'K^+$) approved JLab 
experiment E12-15-008 on $^{40,48}$Ca targets~\cite{Nakamura22}. 
For $F=1$, our calculated $B_\Lambda$ values are close to those calculated 
in Refs.~\cite{IYR17,Byd23}.

\section{Discussion and summary}
\label{sec:sum}

The most significant results of the present phenomenological analysis using 
the $V_\Lambda^{\rm OPT}(\rho)$ methodology are the values at nuclear matter 
density of the two-body and three-body potentials, namely the partial depths 
$D_{\Lambda}^{(2)}$ and $D_{\Lambda}^{(3)}$ as well as their sum $D_{\Lambda}$ 
listed in Eq.~(\ref{eq:total2}). Microscopic models employed in calculations 
of astrophysical scenarios could then be tested for the first time in normal 
nuclear matter by comparing to the present results. SHF models, in contrast, 
do not produce reliable values of $D_{\Lambda}^{(2)}$ and $D_{\Lambda}^{(3)}$. 
This holds for both old~\cite{MDG88} as well as recent~\cite{SHi14,JMNO23} SHF 
calculations. 

\begin{table}
\begin{center}
\begin{tabular}{|c|c|c|c|}
\hline
Model & $D_{\Lambda}^{(2)}$ & $D_{\Lambda}^{(3)}$ & $D_{\Lambda}$ \\
\hline 
Nijmegen ESC16,16$^+$~\cite{ESC16} & $-$43.7 & $+$5.8 & $-$37.9 \\ 
EFT NLO19~\cite{NLO19} & $-$39 to $-$29 & -- & -- \\ 
NLO19 + $\bf{10}$ dominance~\cite{GKW20,Wei24} & $\approx -36$ & 
$\approx +10$ &  \\
EFT N$^2$LO~\cite{N2LO} & $-$33 to $-$38 & -- & -- \\
Femtoscopy~\cite{MHM24} & $-36.3\pm 1.3$(stat.)$^{+2.5}_{-6.2}$(syst.) 
& -- & -- \\ 
\hline 
Hypernuclear constraints (present) & $-37.5 \mp 0.7$ & $9.8 \pm 1.2$ & 
$-27.7 \pm 0.5$ \\  
\hline
\end{tabular}
\caption{Two-body, three-body and total $\Lambda$-nucleus potential depths 
(in MeV) at nuclear matter density $\rho_0=0.17$~fm$^{-3}$ from several 
{\it model} calculations and from hypernuclear binding-energy data.} 
\label{tab:results} 
\end{center} 
\end{table} 

Table~\ref{tab:results} shows comparisons between several recent {\it model} 
predictions and the {\it present} results for $\Lambda$-nuclear potential 
depths. Note that {\it all} model values of $D_{\Lambda}^{(2)}$ are overbound 
with respect to the empirical depth value of $D_{\Lambda}\approx -$30~MeV 
from $\Lambda$ hypernuclei~\cite{GHM16} which tacitly enters such model 
calculations. In comparison with our listed value of $D_{\Lambda}$, this 
overbinding suggests a {\it repulsive} $\Lambda NN$ contribution of depth 
$D_{\Lambda}^{(3)}\approx$~10~MeV, a value indeed reached by one of the EFT 
model calculations~\cite{Wei24} and in agreement with our listed value. 
It would be interesting to repeat this model calculation using our value of 
$D_{\Lambda}=-27.7 \pm 0.5$~MeV as input. The bottom line of this NLO19 EFT 
plus baryon decuplet ($\bf{10}$) dominance model calculation is that hyperons 
are excluded from dense neutron stars owing to a $\Lambda NN$ term of 
a strength commensurate with $D_{\Lambda}^{(3)}\approx$~10~MeV.

\end{document}